\documentclass[aps,apl,longbibliography,superscriptaddress,10pt]{revtex4-1}

\usepackage{graphicx,color}
\usepackage{epsfig}
\usepackage{dcolumn}
\usepackage{amsmath,amssymb,bm}
\usepackage{hyperref,url}
\usepackage[noabbrev]{cleveref}
\usepackage{CJK}
\usepackage{lineno}
\usepackage[english]{babel}
\usepackage{mathptmx} 

\newcommand{\ode}[2]{\frac{d#1}{d#2}}
\newcommand{\oode}[2]{\frac{d^{2}#1}{d#2^{2}}}
\newcommand{\oln}{\bar}

\newcommand\Rey{\text{Re}}  
\newcommand\Str{\text{St}}  
\newcommand\Blc{\text{C}_b} 
\newcommand\etal{\mbox{\textit{et al.}}}

\begin{document}

\title{Vortex shedding in high Reynolds number axisymmetric bluff-body wakes:\\
local linear instability and global bleed control}
\author{A. Sevilla}
\affiliation{\'Area de Mec\'anica de Fluidos, Departamento de
Ingenier\'{\i}a T\'ermica y de Fluidos, Universidad Carlos III de
Madrid.  \mbox{Avda. de la Universidad 30}, 28911, Legan\'es
(Madrid), Spain.}
\author{C. Mart\'{\i}nez-Baz\'an}
\affiliation{\'Area de Mec\'anica de Fluidos, Departamento de
Ingenier\'{\i}a T\'ermica y de Fluidos, Universidad Carlos III de
Madrid.  \mbox{Avda. de la Universidad 30}, 28911, Legan\'es
(Madrid), Spain.}

\begin{abstract}
In the present work we study the large-scale helical vortex shedding regime in the wake of an axisymmetric body with a blunt trailing edge at high Reynolds numbers, both experimentally and by means of local, linear, spatiotemporal stability analysis. In the instability analysis we take into account the detailed downstream evolution of the basic flow behind the body base. The study confirms the existence of a finite region of absolute instability for the first azimuthal number in the near field of the wake. Such instability is believed to trigger the large scale helical vortex shedding downstream of the recirculating zone. Inhibition of vortex shedding is examined by blowing a given flow rate of fluid through the base of the slender body. The extent of the locally absolute region of the flow is calculated as a function of the bleed coefficient, $\Blc=q_b/(\pi R^2u_\infty)$, where $q_b$ is the bleed flow rate, $R$ is the radius of the base and $u_\infty$ is the incident free-stream velocity. It is shown that the basic flow becomes convectively unstable everywhere for a critical value of the bleed coefficient of $\Blc^*\sim 0.13$, such that no self-excited regime is expected for $\Blc>\Blc^*$. In addition, we report experimental results of flow visualizations and hot-wire measurements for increasing values of the bleed coefficient. When a sufficient amount of base bleed is applied, flow visualizations indicate that vortex shedding is suppressed and that the mean flow becomes axisymmetric. The critical bleed coefficient predicted by linear instability analysis is shown to fall within the experimental values in the range of Reynolds numbers analyzed here.
\end{abstract}

\maketitle

\section{Introduction}

The phenomenon of vortex shedding in bluff-body wakes has been successively addressed in the past with the aid of hydrodynamic instability theory. In the case of two-dimensional wakes, starting with the pioneering work of Mathis \emph{et al.}~\cite{mathis84}, many experimental and numerical studies~\cite{provansal87,jackson87,hanne89,karnia89,schumm94,legal00} have shown that von-K\'arm\'an vortex street is the consequence of a supercritical Hopf bifurcation. In this context a global linear stability analysis of the flow provides the critical value of the Reynolds number, as well as the complex frequency as an eigenvalue of the problem~\cite{jackson87}. On the other hand, the use of local stability analysis to study this type of flows, assuming a parallel basic flow at each downstream station, has the advantage of requiring very little computational cost, but has the disadvantage of needing additional criteria to predict frequencies associated with the growth of the unstable global mode.

Different \emph{global frequency selection criteria}, whose degree of success depends on the flow under study, have been proposed in the past, i.e. Koch's hydrodynamic resonance criterion~\cite{koch85} (HR), Pierrehumbert's maximum growth criterion~\cite{pierrehumbert84} (MG), and Monkewitz and Nguyen's initial growth criterion~\cite{monk87} (IG). Later on, the study of \emph{linear global} instability of \emph{weakly non-parallel} shear flows~\cite{monk93} provided with the first criterion based on solid theoretical grounds, commonly referred to as Chomaz--Huerre--Redekopp criterion~\cite{chr91} (CHR). Such studies showed that the existence of a sufficiently long region of local absolute instability is a necessary condition for the flow to sustain a linearly unstable global mode. However, more recent work on the theory of \emph{non-linear} instability of \emph{slowly divergent} flows (see Pier and Huerre~\cite{pier01} and references therein) has revealed that, in the case of two-dimensional wake-like flows~\cite{pier01},  a bifurcation to self-sustained oscillations is triggered whenever a region of local absolute instability exists in the flow. In addition, in the case of wakes, the non-linear global mode has been found to be a \emph{steep} or \emph{elephant} mode  whose frequency, in coincidence with the aforementioned IG criterion, is imposed by the first absolutely unstable downstream station.

As was already pointed out by Pier~\cite{pier02}, the near wake of bluff bodies is far from being slender due to the presence of the recirculating bubble. Thus, a parallel-flow-like instability analysis of the near wake is not a rigorous approach to the problem. Nevertheless, in the last years several studies~\cite{hanne89,hammond97,pier02} showed that a local, spatiotemporal, linear instability analysis performed as a function of the downstream position can give good global predictions when the appropriate frequency selection criterion is chosen.

Although vortex shedding behind axisymmetric bluff bodies has received less attention than its two-dimensional counterpart, the wake behind a sphere has been studied in detail in the past. This kind of flow serves as a prototype for a more general class of axisymmetric wakes with a toroidal recirculating region close to the body. The self-excited vortex shedding regime does not appear only in the laminar wake at low or moderate values of the Reynolds number, but it also persists in the turbulent wake as large-scale structure. In the case of the turbulent wake behind a sphere, early flow visualizations performed by Taneda~\cite{taneda78} and hot wire measurements by Achenbach~\cite{achen74} showed the existence of vortex shedding as a coherent phenomenon superimposed to the turbulent flow field. This structure can be explained as a superposition of equal-strength, counter-rotating helical modes with azimuthal numbers $m=\pm 1$. Later on Kim and Durbin~\cite{kim88} showed that the turbulent self-excited regime is characterized by a Strouhal number almost independent of the Reynolds number whose value, typical of vortex shedding phenomena, was $\Str\sim 0.2$. They also discovered that the shedding mode was insensitive to low levels of external acoustic excitation for forcing frequencies far enough from the natural one, a feature typical of globally unstable flows.

Previous works based on configurations similar to the one under study here have focused on the following issues. Monkewitz~\cite{monk88c} studied the viscous and inviscid instability of generic, natural wake velocity profiles parameterized by a velocity ratio, $\Lambda$, and a momentum thickness parameter, $N$, without a detailed specification of their downstream evolution for a specific geometry. He concluded that helical vortex shedding in the wake of axisymmetric objects at high Reynolds numbers is related to the presence of local absolute instability in the near wake. Schwarz \emph{et al}.~\cite{schwarz94} performed a direct numerical simulation of the wake of a bullet-shaped body at two different Reynolds numbers, namely $\Rey=500$ and $\Rey=10^3$. They found that, as happens in the case of the wake behind a sphere, the wake is dominated by global instability modes of azimuthal number $m=\pm 1$ for both Reynolds numbers. Thus, it seems that the flows characterized by  an axisymmetric wake with a recirculating zone in the near field share the same main feature, i.e. the dominant instability mode, which triggers large-scale structure in the wake, is the first azimuthal mode.

The control of two-dimensional wakes has also been extensively studied in the past. Different control strategies~\cite{wood67,bearman67,schumm94}, like bleed, suction, heating, body vibration and body rotation among others, have been successfully used to inhibit vortex shedding occurring behind two-dimensional  bodies. In particular, Bearman~\cite{bearman67}, and more recently Schumm \etal~\cite{schumm94}, found that a
sufficient amount of base bleed could completely suppress vortex shedding in the turbulent wake of a blunt-edged two-dimensional body. Similarly, Leu and Ho~\cite{leu00} showed that the K\'arm\'an vortex street was also suppressed when a suction coefficient greater than a critical one was applied, $U_s/U_\infty \approx 0.5$. In the case of axisymmetric bodies, Weickgenannt and Monkewitz~\cite{weick00} studied the control of a bullet-shaped body wake by a rear-mounted disk. To our knowledge, base bleed control of axisymmetric wakes has not been studied and our main purpose in this paper is to examine this wake control mechanism in an axisymmetric body.

Thus, the present paper studies the helical vortex shedding regime in the wake of a slender body of revolution with a blunt trailing edge and investigates the inhibition of vortex shedding by means of base bleed. The work includes a linear stability analysis as well as experimental results obtained from flow visualizations and hot-wire measurements. The stability calculations are performed at a fixed Reynolds number of $3\times 10^3$, with a bleed coefficient, $\Blc=u_b/u_\infty$, where $u_b$ is the bleed velocity and $u_\infty$ is the free stream velocity, which varied from $0$ to $0.2$. In order o determine the velocity profiles of the basic flow we decided to perform a detailed computation of flow field rather than to use parametric basic velocity profiles. This allowed us to unambiguously obtain the information required to evaluate the local absolute growth rates and frequencies in our particular geometry. The basic flow is obtained as a steady, laminar, axisymmetric solution of the incompressible Navier-Stokes equations. Although at $\Rey=3\times 10^3$ the flow is unstable, the solution can be interpreted as the pseudo-steady flow obtained in fully three-dimensional simulations just before the beginning of unsteadiness~\cite{hanne89}. Furthermore, we provide with predictions of the global instability properties of the flow obtained with different frequency selection criteria, as well as a critical value of the bleed coefficient for which absolute instability is completely inhibited, $\Blc^*$. Additionally, flow visualizations are reported in the range of Reynolds numbers $500<\Rey<2.8\times 10^3$, and hot-wire measurements in the range $4\times 10^3<\Rey<1.2\times 10^4$.

The paper is structured as follows: the flow configuration under study, as well as the formulation and results obtained for both the basic flow and the local instability calculations, are presented in section II. The experimental configuration and results are discussed in section III, and we finish with conclusions in section IV.

\section{Linear instability analysis}

In this section, we present the technique employed to obtain the basic flow field, and the linearized disturbance evolution equations are briefly discussed.

\subsection{Flow configuration}\label{config}

\begin{figure}[!h]
\begin{center}\includegraphics[width=0.5\textwidth]{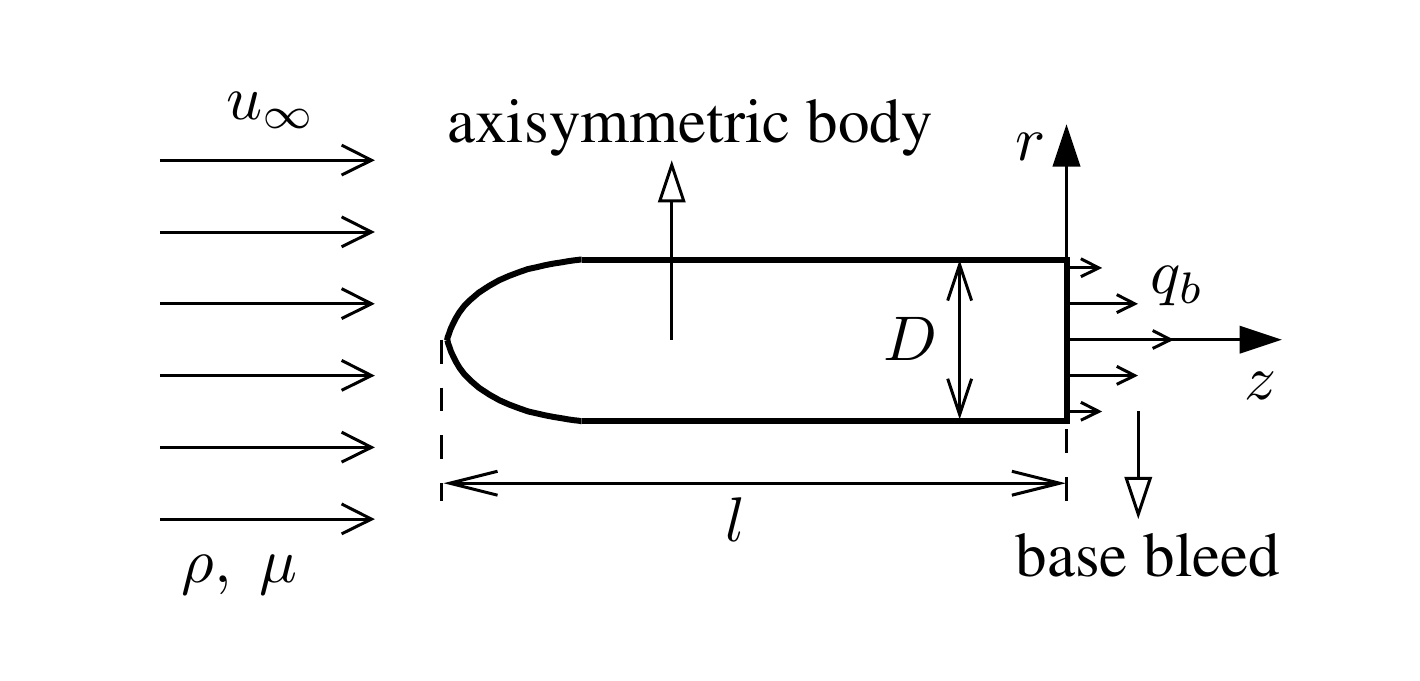}\end{center}
\caption{Scheme of the flow configuration under study.}\label{fig1}
\end{figure}

The flow configuration under study here, and shown in Fig.~\ref{fig1}, corresponds to a slender body of revolution with a 3:1 aspect ratio ellipsoidal nose, total length $l$ and a blunt trailing edge of diameter $D$ placed into a uniform stream of fluid of density $\rho$ and viscosity $\mu$ at zero angle of attack. The free-stream velocity is $u_\infty$, and the Mach number, $\text{Ma}=u_\infty/c_0\ll 1$, where $c_0$ denotes the sound speed of the ambient medium. A flow rate $q_b$ of the same fluid is blown through the base with a given axial velocity profile $u_b(r),\;0\leq r\leq R=D/2$. Thus, a dimensionless bleed coefficient can be defined as $\Blc=m_b/m_\infty$, where $m_b=2\pi \int_0^R \rho \, u_b(r) \, r \, dr$ and $m_\infty=\pi \, R^2 \, \rho \, u_{\infty}$ denote the bleed and incident mass flow rates respectively, the latter based on the body-base area and outer velocity. In the case of uniform bleed profile considered here the bleed coefficient simplifies to the bleed to free-stream velocity ratio, $\Blc=u_b/u_\infty$.

For a given nose geometry, there are three dimensionless parameters governing the problem, namely the Reynolds number, $\Rey=\rho\,u_\infty D/\mu$, the bleed coefficient, $\Blc$, and the body length to diameter ratio, $\text{L}=l/D$. The product of the Reynolds number and $\text{L}$ is the Reynolds number based on the total body length, $\Rey_l=\rho\,u_\infty l/\mu$. Provided that the surface roughness is sufficiently small to be negligible, the parameter $\Rey_l$ determines whether the boundary layer developing on the surface of the slender body remains laminar throughout the whole body length or, otherwise, becomes turbulent prior to separation. For a geometry similar to the one under consideration here, Weickgenannt and Monkewitz~\cite{weick00} found that the boundary layer remained laminar at the trailing edge for  $\Rey\lesssim 2\times 10^4$. In our case, the experiments were performed with a scaled model with $\text{L}= 9.8$. Thus, the estimated Reynolds number for transition to turbulence in our body's boundary layer is $\Rey\approx 1.2\times 10^4$, a value close to the maximum value considered in the present work. Therefore, in the rest of the paper, we will assume that the boundary layer remains laminar at the trailing edge.

Following the preceding considerations, dimensional analysis implies that the shedding Strouhal number, $\Str=fD/u_\infty$ where $f$ denotes the vortex shedding frequency, has the following functional dependence: $\Str=\Str\left(\Rey,\Blc,\text{L}\right)$. In addition, the dependence of the critical bleed coefficient can be written as  $\Blc^*=\Blc^*\left(\Rey,\text{L}\right)$. Since, in our study, the parameter $\text{L}$ was kept constant and equal to 9.8, we will mainly focus on the characterization of $\Str=\Str\left(\Rey,\Blc\right)$ and $\Blc^*=\Blc^*\left(\Rey\right)$.

\subsection{Basic flow calculation}\label{profile}

Cylindrical coordinates will be denoted $\left(z,r,\phi\right)$, standing respectively for the axial, radial and azimuthal coordinates. The origin of the reference frame is located at the center of the body base. The velocity field is introduced as $\bm{v}=\left(u,v,w\right)$ where $u$, $v$ and $w$ denote the axial, radial and azimuthal velocity components, and the pressure field is named $p$.

Under the assumptions made in section \ref{config}, the flow is governed by the incompressible Navier-Stokes equations, which can be written in dimensionless form as,
\begin{subequations}
 \label{conservation}
 \begin{eqnarray}
  \nabla\cdot\bm{v}&=&0\,,\label{consa}\\
  \frac{\partial\bm{v}}{\partial \tau}+\bm{v}\cdot\nabla\bm{v}+\nabla p&=&\Rey^{-1}\triangle\bm{v}\,,\label{consb}
 \end{eqnarray}
\end{subequations}
together with the appropriate boundary conditions. All quantities have been made nondimensional with length, velocity and pressure scales $D$, $u_\infty$ and $\rho u_\infty^2$. The dimensionless axial and radial coordinates will be denoted $Z=z/D$ and $\eta=r/D$, and the dimensionless time will be named $\tau=t\,u_\infty/D$, where $t$ denotes the dimensional time.

The decomposition of the flow into a basic steady state and small amplitude disturbances will be introduced as follows,
\begin{equation}
\left(\bm{v},p\right)=(\bm{V}+\hat{\bm{v}},P+\hat{p})\,,\label{perturbations}
\end{equation}
where $\bm{V}=\left(V_z(Z,\,\eta),V_r(Z,\,\eta)\right)$ and $P(Z,\,\eta)$ constitute the basic, axisymmetric, steady flow and $\hat{\bm{v}}$ and $\hat{p}$ represent the unsteady disturbance velocity and pressure fields respectively. The basic flow is enforced to satisfy Eqs.~eq\ref{conservation} in their steady axisymmetric form,
\begin{subequations}
 \label{conservation2}
 \begin{eqnarray}
  \nabla\cdot\bm{V}&=&0\,,\label{cons2a}\\
  \bm{V}\cdot\nabla\bm{V}+\nabla P&=&\Rey^{-1}\triangle\bm{V}\,.\label{cons2b}
 \end{eqnarray}
\end{subequations}
To determine the basic velocity profiles required to perform the local instability analysis, Eqs.~\eqref{conservation2} were solved by means of a standard finite volume numerical method using a staggered grid, and the classical SIMPLE algorithm for pressure correction. 

\begin{figure}[!h]
\begin{center}\includegraphics[width=0.4\textwidth]{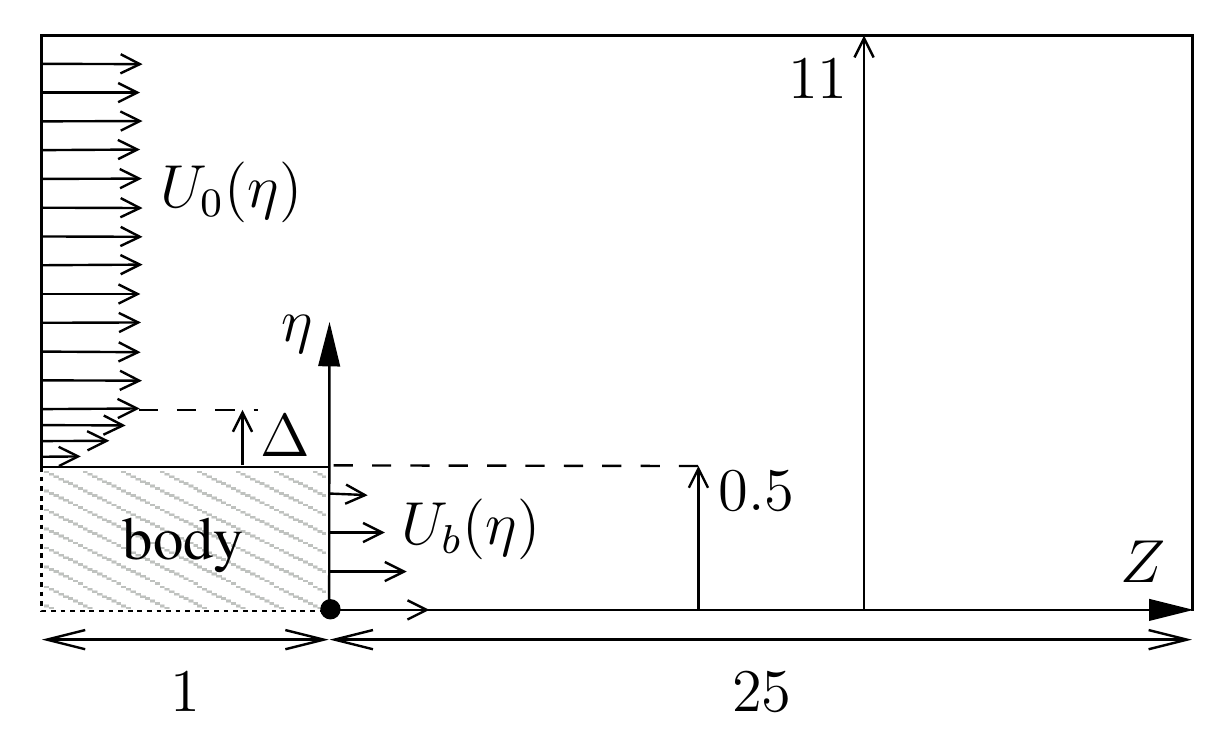}\end{center}
\caption{Computational domain used for the calculation of the basic flow.}\label{fig2}
\end{figure}

The computational domain used is shown in Fig.~\ref{fig2}. It extends one diameter upstream and $25$ diameters downstream from the base, and has a cross-stream extent of $11$ diameters. The numerical results obtained were checked to be independent of mesh refinement. The boundary conditions for Eqs.~\eqref{conservation2} are the following,
\begin{subequations}
 \label{boundary}
 \begin{eqnarray}
 1/2\leq\eta\leq 11,\,Z=-1&:&\;V_z=U_0(\eta),\;V_r=P=0, \label{bouna}\\
 \eta=1/2,\,-1\leq Z\leq 0&:&\;V_z=V_r=0, \label{bounb}\\
 0\leq\eta\leq 1/2,\,Z=0&:&\;V_z=U_b(\eta),\;V_r=0, \label{bounc}\\
 \eta=11,\,-1\leq Z\leq 25&:&\;V_r=\partial V_z/\,\partial\eta=0, \label{bound}\\
 \eta=0,\,0\leq Z\leq 25&:&\;V_r=\partial V_z/\,\partial\eta=0, \label{boune}\\
 0\leq\eta\leq 11,\,Z=25&:&\;P=V_r=\partial V_z/\,\partial Z=0\,,\label{bounf}
   \end{eqnarray}
\end{subequations}
where $U_0(\eta)$ is the incoming free-stream velocity profile whose boundary layer thickness is $\Delta$ and $U_b(\eta)$ is the bleed velocity profile. For the incident stream, $U_0$, we decided to use a parabolic boundary layer profile of dimensionless thickness $\Delta = 0.05$, matched to the uniform free stream velocity $u_\infty$. Although for the bleed profile, $U_b$, we used both a uniform and a parabolic profile, the resulting basic flows were almost identical and, therefore, we only report here the results obtained with the uniform bleed velocity profile.

At moderately high, supercritical Reynolds numbers, the numerical flow field remains steady, probably because unsteadiness begins with a symmetry-breaking bifurcation, while in our calculations we impose axisymmetry. Therefore, the computed flow is equivalent to the pseudo-steady flow present in supercritical, unsteady calculations just before the onset of time dependence. Although some authors~\cite{karnia89} have proposed the mean flow of the developed, unsteady flow field to be the right choice for linear stability calculations of supercritical regimes, in this case it would require to perform a fully three-dimensional, unsteady numerical simulation which is out of the scope of the present study. Thus, we will follow Hannemann and Oertel~\cite{hanne89} selection, who successfully used the pseudo-steady flow in a linear instability analysis of a flat plate wake.

\begin{figure}[!h]
\begin{center}\includegraphics[width=0.5\textwidth]{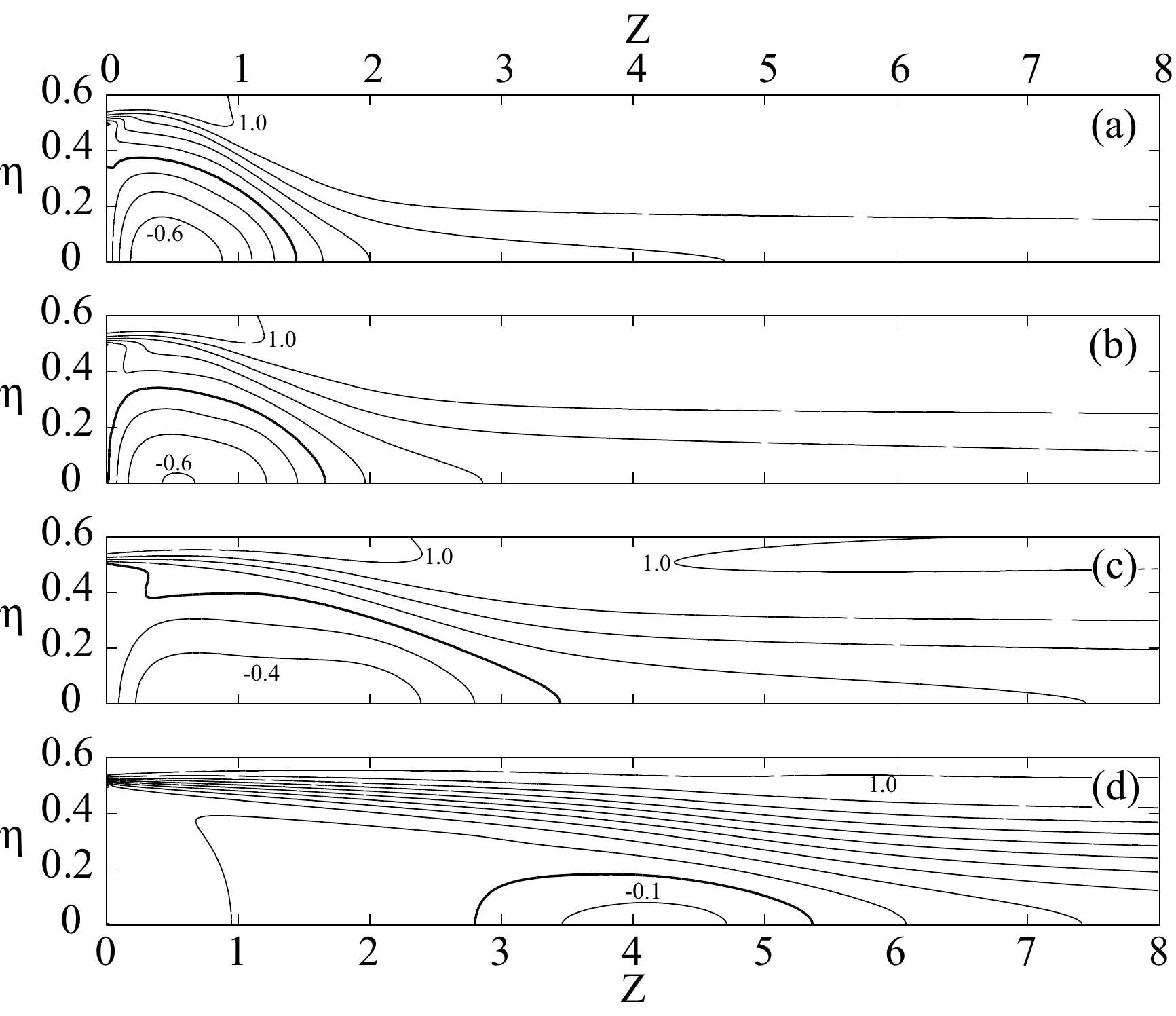}\end{center}
\caption{Contours of constant basic-flow axial velocity for different values of the bleed coefficient: (a) $\Blc=0$, (b) $\Blc =0.05$, (c) $\Blc=0.1$  and (d) $\Blc=0.14$.}\label{fig3}
\end{figure}

Figure~\ref{fig3} shows the contours of constant axial velocity for $\Rey=3\times 10^{3}$ and for several values of the bleed coefficient, namely $\Blc=0,\,0.05,\,0.1$ and $0.14$. It can be observed that as the bleed coefficient increases it affects the near wake in two main aspects:  on the one hand, the length of the recirculating region increases and, on the other hand, the radial gradients of the axial velocity component decrease in all the downstream positions. Notice that in the natural wake case, $\Blc=0$, the length of the recirculating bubble is about one base diameter, and the magnitude of reverse flow can be as large as $50\%$ of the free-stream velocity. On the contrary, for a base bleed of $0.1$ (Fig.~\ref{fig3}c), the recirculating region is bounded by two free-stagnation points and has a length of approximately 3 diameters, while the maximum reverse flow is less than $25\%$ of the free-stream value. It is well known~\cite{monk88c} that reverse flow promotes local absolute instabilities, and that the absolute growth rate decreases as the velocity gradient decreases. Thus, one can expect the local absolute instability  to be completely inhibited for a sufficiently large base bleed.

In the following subsection we will briefly introduce the formulation of the local instability problem.

\subsection{Stability analysis}

Substituting Eq.~\eqref{perturbations} into Eqs.~\eqref{conservation}, and neglecting quadratic terms in the perturbations, we obtain the following linear evolution equations for the disturbance fields,
\begin{subequations}
 \label{conservation3}
 \begin{eqnarray}
  \nabla\cdot\hat{\bm{v}}&=&0,\label{cons3a} \\
  \frac{\partial\hat{\bm{v}}}{\partial\tau}+\bm{V}\cdot\nabla\hat{\bm{v}}+\hat{\bm{v}}\cdot\nabla\bm{V}+\nabla \hat{p}&=&\Rey^{-1}\triangle\hat{\bm{v}}\,,\label{cons3b}
 \end{eqnarray}
\end{subequations}
which are four linear partial differential equations with four unknowns $\left(\hat{u},\hat{v},\hat{w},\hat{p}\right)$. Notice that Eqs.~\eqref{conservation3} are only homogeneous in time and in the azimuthal coordinate, but not in the axial and radial coordinates, and only a \emph{global instability analysis} would be possible up to this point~\cite{jackson87}. However \emph{local instability analysis} simplifies Eqs.~\eqref{conservation3} by assuming a basic parallel shear flow with the following form,
\begin{equation}
\bm{V}=\left(U(\eta),0,0\right),\label{local}
\end{equation}
where the radial velocity component, as well as the streamwise variation of the axial velocity component, are both neglected in the stability analysis. Under the assumptions made in the local approximation, the general solution of Eqs.~\eqref{conservation3} is the superposition of travelling-wave solutions, which are normal modes with harmonic dependence on the variables $\left(Z,\phi,\tau\right)$,
\begin{equation}
\left(\hat{u},\hat{v},\hat{w},\hat{p}\right)=\left(\oln{u}(\eta),\oln{v}(\eta),\oln{w}(\eta),\oln{p}(\eta)\right)e^{i\left(KZ+m\phi-\Omega\tau\right)}.
\label{harmonic}
\end{equation}
In Eq.~\eqref{harmonic} the unknowns are decomposed into an $\eta$-dependent complex amplitude $\left(\oln{u},\oln{v},\oln{w},\oln{p}\right)$, and a complex exponential function containing the $\left(Z,\phi,\tau\right)$ dependence. Notice that physical magnitudes can be obtained only after taking the real part of the corresponding complex expression. Real and imaginary parts of complex numbers will hereafter be denoted by $r$- and $i$-subscripts respectively. In Eq.~\eqref{harmonic}, $K=K_r+iK_i$ is the complex perturbation wavenumber, $m=0,\pm 1,\pm 2,\cdots$ stands for the azimuthal mode and $\Omega=\Omega_r+i\Omega_i$ refers to the perturbation frequency, related to the Strouhal number simply by $\Str=\Omega_r/(2\pi)$. In the inviscid limit, $\Rey\rightarrow\infty$, it is possible to reduce the system of equations~\eqref{conservation3} to a single ordinary differential equation for the pressure disturbance amplitude $\oln{p}$, given by~\cite{michal82,gord01}
\begin{equation}
\oode{\oln{p}}{\eta}+\left(\frac{1}{\eta}-\frac{2 \, K}{K
\,U-\Omega} \, \, \ode{U}{\eta}\right) \,
\ode{\oln{p}}{\eta}-\left(K^{2}+\frac{m^{2}}{\eta^{2}}\right) \,
\oln{p}=0\,, \label{ode}
\end{equation}
with the following boundary conditions
\begin{equation}
\left\{\begin{array}{ll}
    \eta=0 & \oln{p}\neq \infty\\
    \eta\rightarrow \infty & \oln{p}\rightarrow 0\,.
    \end{array}
\right. \label{bc2}
\end{equation}
Equations~\eqref{ode}-\eqref{bc2} are the eigenvalue problem to be solved in order to obtain non-trivial eigenfunctions $\left(\oln{u},\oln{v},\oln{w},\oln{p}\right)$ with associated eigenvalues ($K,\Omega$), the latter usually expressed as solutions of a \emph{local dispersion relation} $D\left(K,\Omega;\,m,Z,\Rey,\Blc\right)=0$. In the last expression it is explicitly shown the parametric dependence of the eigenvalue problem on the azimuthal number, the downstream position, the Reynolds number and the bleed coefficient. Since our calculations will be performed at a fixed Reynolds number of $3\times 10^3$, the results presented in this paper will be based on the inviscid instability equations~\eqref{ode})-\eqref{bc2}. Moreover, the parametric dependence of the results on the Reynolds number will not be taken into consideration, since its effect is expected to be small as long as $\Rey$ is sufficiently large. In the context of spatiotemporal instability analysis, both the frequency and the wavenumber are in general complex, and the absolute or convective nature of the local instability can be established by examining the solutions of the dispersion relation with zero group velocity, $d\Omega/dK(K^{0)})=0$, that satisfies the Briggs-Bers criterion~\cite{huerre00}. In the complex wavenumber plane, the latter condition is equivalent to the existence of a saddle-point in the dispersion relation at $K^{(0)}$, $dD/dK(K^{(0)})=0$. The values of wavenumber and frequency at the saddlepoint, $(K^{(0)},\Omega^{(0)})$, are usually called \emph{absolute wavenumber} and \emph{absolute frequency}, respectively. Of particular importance to predict the behavior of the flow is the \emph{absolute growth rate} defined as the imaginary part of the frequency at the saddle-point, $\Omega_i^{(0)}$. If $\Omega_i^{(0)}(Z)>0$ ($\Omega_i^{(0)}(Z)<0$), the flow is locally absolutely (convectively) unstable at location $Z$. The real part of the absolute frequency, $\Omega_r^{(0)}(Z)$, can not be immediately interpreted as the frequency associated with global instability, since its value depends on the downstream position inside the wake. As already stated before, a frequency selection criterion has to be used to determine the downstream position which describes the global instability properties. We may introduce the \emph{local Strouhal number} as $\Str^{(0)}(Z)=\Omega_r^{(0)}/(2\pi)$. In addition, the real value of the absolute wavenumber, $K_r^{(0)}(Z)$, can be related to a \emph{local wavelength}, $\lambda^{(0)}(Z)=2\pi/ K_r^{(0)}$. Details of the numerical method used to solve the eigenvalue problem~\eqref{ode}-\eqref{bc2} are given elsewhere~\cite{sevilla02}.

\subsection{Results}

The procedure we followed in the calculations consisted of solving a local stability problem at any given downstream location, for fixed values of both the Reynolds number and the bleed coefficient. The Reynolds number was $3\times 10^{3}$ in all the calculations, while the bleed coefficient was varied from $\Blc=0$ to $\Blc=0.14$. Since the different downstream positions are decoupled from one to another in this simple analysis, the only input needed to start the calculations is the radial profile of the axial velocity component, $U(\eta)$, obtained as described in section~\ref{profile}.\\

\begin{figure}[!h]
\begin{center}\includegraphics[width=0.45\textwidth]{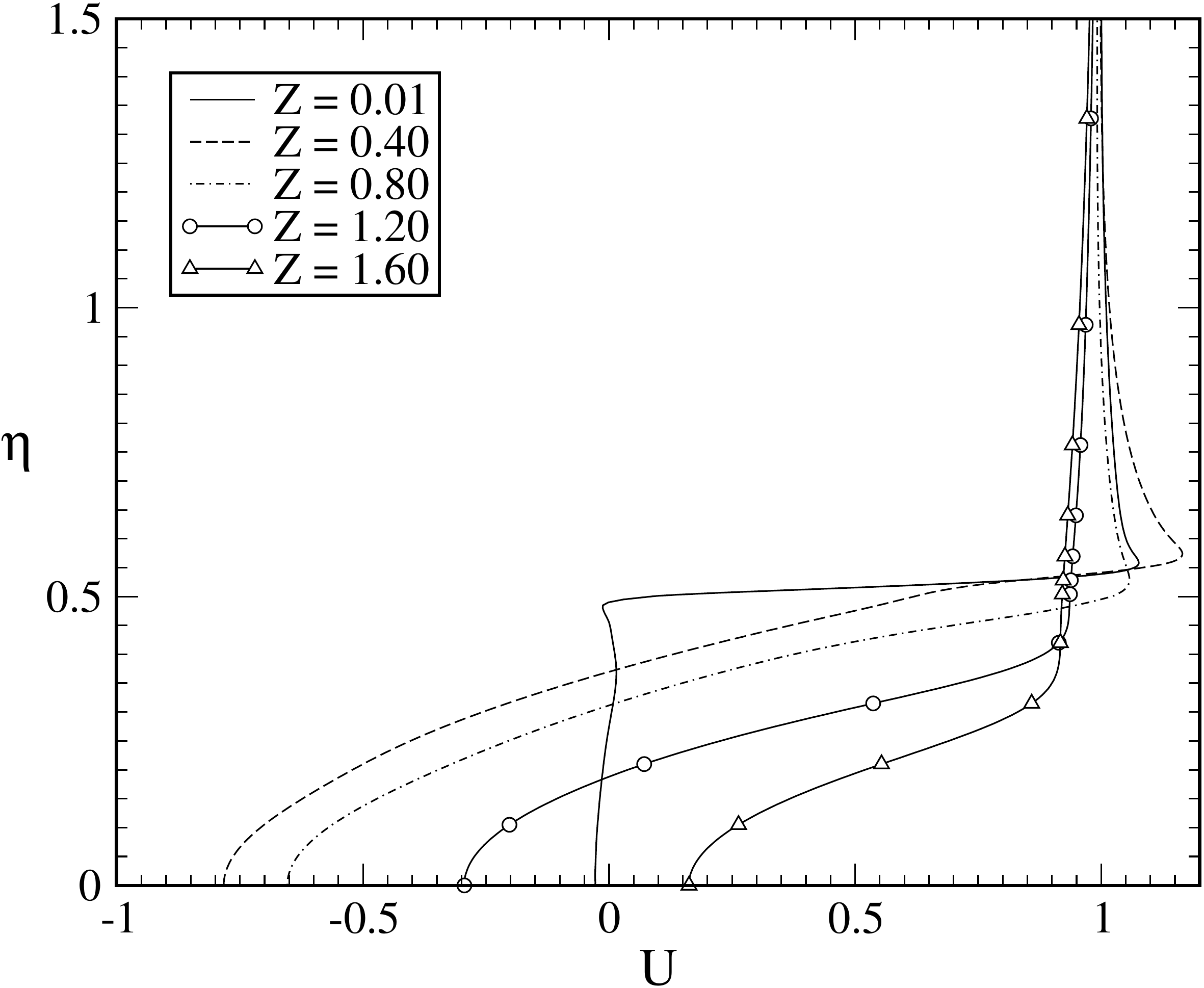}\end{center}
\caption{Computed radial profiles of dimensionless axial velocity at several downstream positions. $\text{Re}=3000$ and $\text{C}_b=0$.}\label{fig4}
\end{figure}

\begin{figure}[!h]
\begin{center}\includegraphics[width=0.45\textwidth]{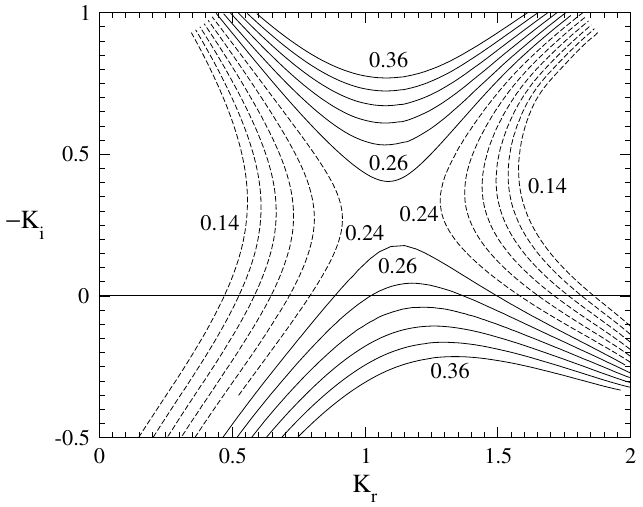}\end{center}
\caption{Spatial instability branches for $\text{C}_b=0$, $Z=0.5$, $|m|=1$.}\label{fig5}
\end{figure}

Figure~\ref{fig4} shows several near-wake velocity profiles for $\Blc=0$. The spatial instability branches in the complex wavenumber plane for the same values of the control parameters and at $Z=0.5$, a location inside the recirculating bubble, are shown in Fig.~\ref{fig5} for the $|m|=1$ azimuthal mode. Each curve corresponds to a constant value of $\Omega_i$, indicated in the plot,  and varying values of $\Omega_r$. Notice the existence of a saddle-point which fulfills the Briggs--Bers criterion, since the spatial branches involved in the saddle-point separate two different halves of the complex wavenumber plane for high enough values of $\Omega_i$. It can also be observed that under the conditions of Fig.~\ref{fig5} the flow is locally, absolutely unstable, since $\Omega_i(K^{(0)})>0$. We performed the same type of calculations in the case of the axisymmetric mode, $m=0$, giving a convective instability. Thus, since the $|m|=1$ mode is the dominant one in the flow configuration under study, we will only consider the first azimuthal mode hereafter in the paper. 

\begin{figure}[!h]
\begin{center}\includegraphics[width=0.55\textwidth]{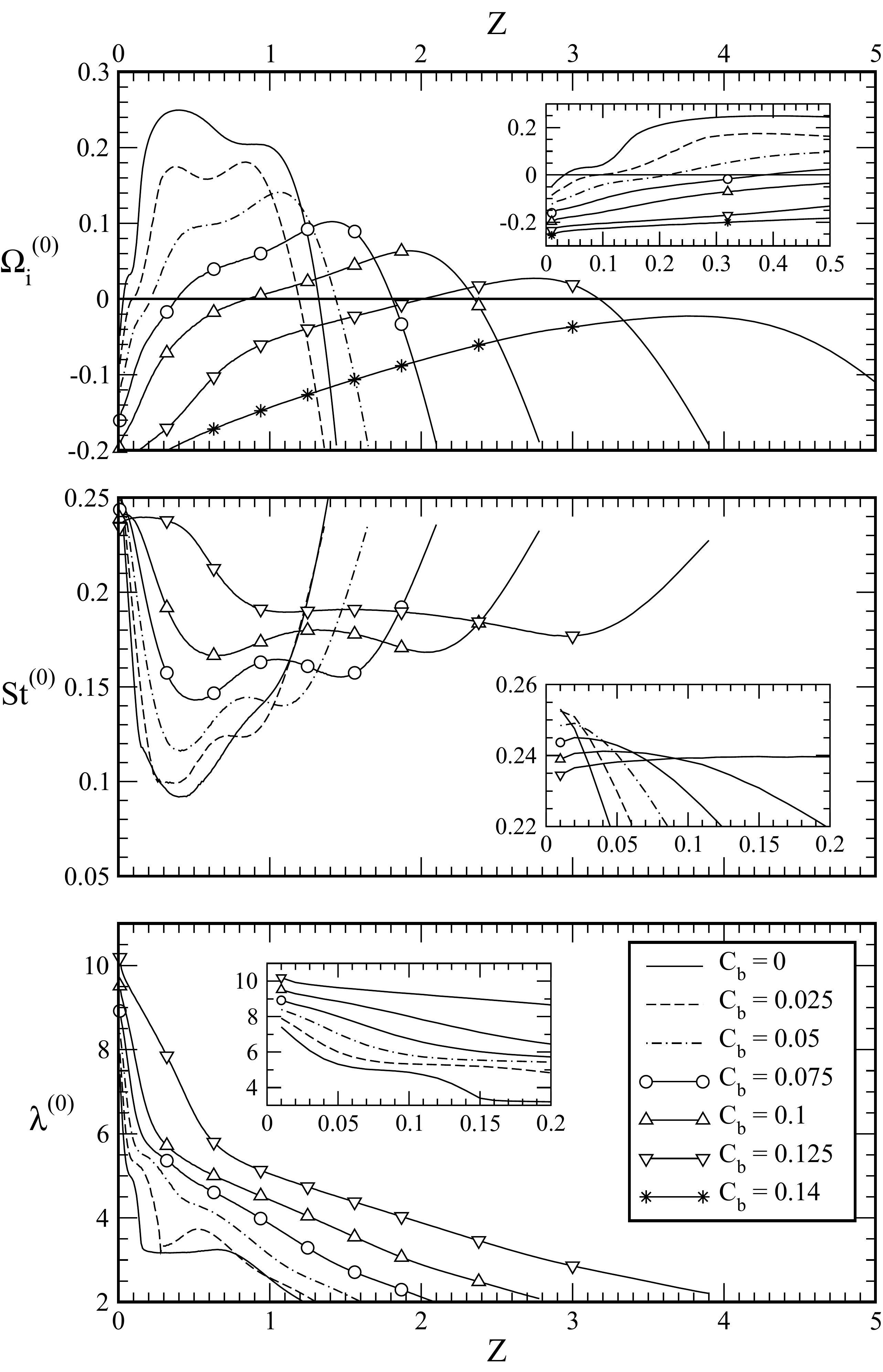}\end{center}
\caption{Absolute growth rate, $\Omega_i^{(0)}$, Strouhal number, $\text{St}^{(0)}$, and wavelength, $\lambda^{(0)}$, for the first azimuthal mode as a function of downstream distance $Z$ for several values of the bleed coefficient: -----,\;$\text{C}_b=0$\,;\;--\,--\,--,\;$\text{C}_b=0.025$\,;\;--\,$\cdot$\,--,\;$\text{C}_b=0.05$\,;\;--$\circ$--,\;$\text{C}_b=0.075$\,;\;--$\vartriangle$--,\;$\text{C}_b=0.1$\,;\;--$\triangledown$--,\;$\text{C}_b=0.125$ \,;\;--$\ast$--,\;$\text{C}_b=0.14$.}\label{fig6}
\end{figure}

Once the eigenvalues $\left(\Omega,K\right)$ have been calculated at two different positions $Z_1,\,Z_2$ such that $|\,\Delta Z\,|=|Z_2-Z_1|\ll 1$, a first guess for the eigenvalues at a different station $Z_3$ such that $|\,Z_3-Z_2\,|\ll 1$ can be chosen by linear extrapolation, $\Omega(Z_3)\approx 2\,\Omega(Z_2)-\Omega(Z_1)$. This estimation provides a good starting point for the Newton-Raphson iterative scheme, and also improves convergence speed. The same procedure has been implemented for all the reported values of the bleed coefficient. The results of these computations can be observed in Fig.~\ref{fig6}, where it is shown the downstream evolution of the local absolute growth rate, $\Omega^{(0)}_i$, Strouhal number, $\Str^{(0)}$ and wavelength, $\lambda^{(0)}$ for the $|m|=1$ mode and several values of the bleed coefficient $\Blc$.

Let us begin by considering the case without base bleed, which corresponds to the solid line without symbols. The absolute growth rate is negative at the surface of the body, and rapidly becomes positive at a very short distance from the base, indicating the transition from local convective to local absolute instability. The IG criterion, coincident with the steep non-linear global mode criterion as already stated in the introduction, applies precisely at this transition point, denoted hereafter by $Z_{IG}$. As $Z$ increases past $Z_{IG}$, the AGR increases, until a maximum value is reached at a position $Z_{MG}$, postulated by the MG criterion as the station dominating the global behavior of the wake. The AGR starts decreasing from $Z_{MG}$ until it eventually becomes zero at the location $Z_{HR}$, where Koch's criterion applies. For $Z>Z_{HR}$ the AGR remains negative, showing that the region of local absolute instability is of finite extent, corresponding to stations $Z_{IG}<Z<Z_{HR}$. Notice that this region approximately coincides with the recirculating bubble, indicating a clear correspondence between reverse flow and absolute instability~\cite{pier02}. The same general picture can be observed for increasing values of the bleed coefficient. As $\Blc$ increases the stations $Z_{IG},\,Z_{MG},\,Z_{HR}$ move to higher values of $Z$ indicating that the recirculating region also moves downstream. At the same time, the maximum value of the AGR throughout the whole wake, $\Omega_i^{(0)}(Z_{MG})$, decreases, since the maximum shear also decreases with $\Blc$. Thus, the bleed mechanism provides two competing effects in order to inhibit the global instability: a larger absolutely unstable region implies more risk of global instability while, on the other hand, a decreasing value of  maximum AGR reduces the possibility of triggering global instability. If suction, instead of blowing, was applied at the base, this situation would be reversed~\cite{leu00}: the maximum AGR would increase while the length of the absolutely unstable region would decrease. However, notice that the flow is more slender as $\Blc$ increases and, contrary to the case of base suction, the parallel-flow assumption becomes more rigorous. This feature has its counterpart in the plots of Fig.~\ref{fig6}, showing that the streamwise derivatives of the local instability properties decrease as $\Blc$ increases, which indicates that the local analysis performed in this work becomes a better first-order approximation to the global instability analysis. Although suppression of global modes could also be achieved for a certain value of the suction coefficient, the main objective of this section is to show that complete inhibition of local absolute instability can be achieved at a critical value of the bleed coefficient, $\Blc^*$. Based on the results of Fig.~\ref{fig6}a, we can conclude that the critical bleed coefficient results $0.125<\Blc^*<0.14$.

\begin{figure}[!h]
\begin{center}\includegraphics[width=0.5\textwidth]{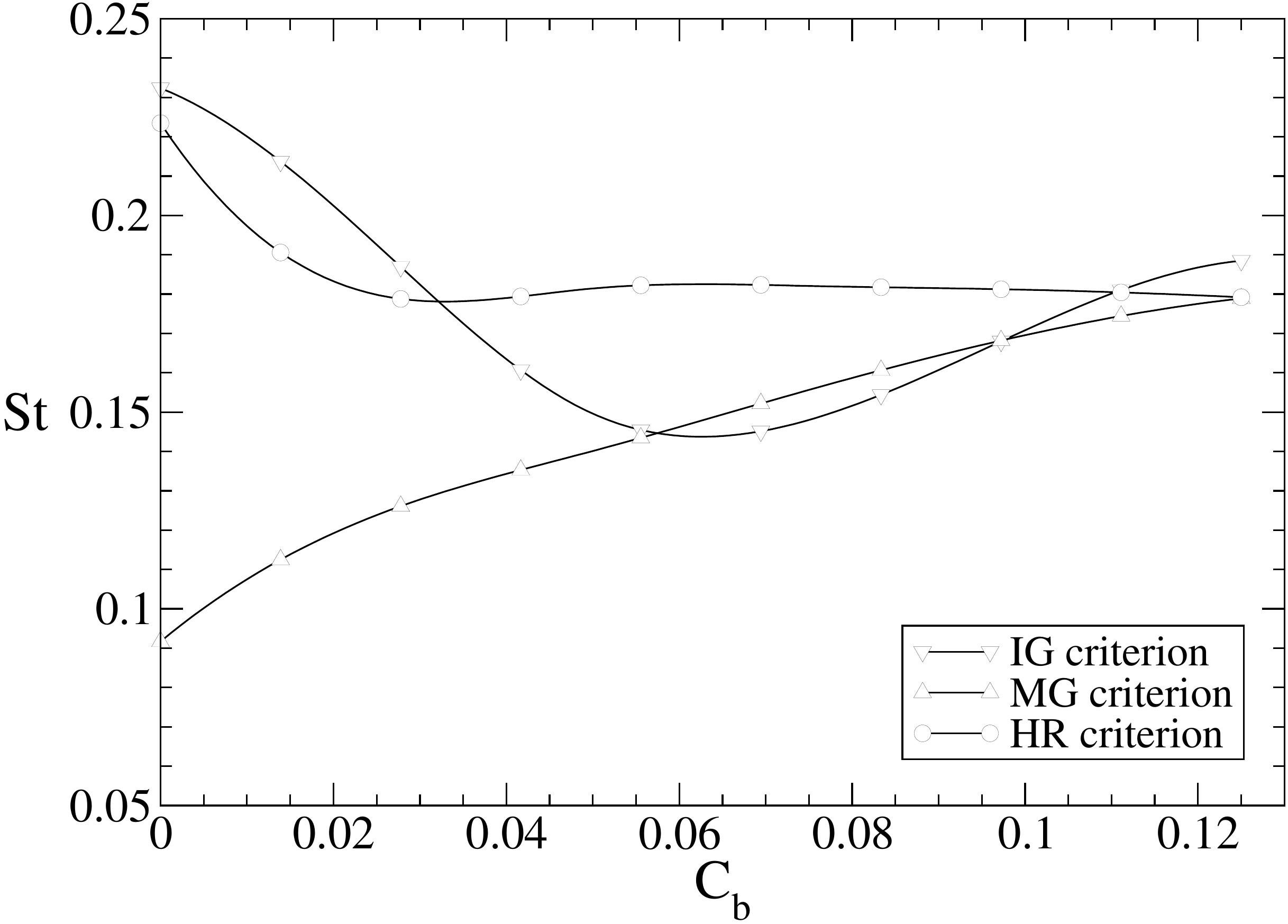}\end{center}
\caption{Global Strouhal number given by the IG, MG and HR frequency selection criteria, as a function of
the bleed coefficient.}\label{fig7}
\end{figure}

The IG (or steep global), MG and HR  criteria have been applied to provide specific predictions for the vortex shedding frequency. Therefore, a global Strouhal number, $\Str$, associated with each frequency selection criterion has been defined. For instance, the global Strouhal number related to the IG criterion is simply given by $\Str=\Str^{(0)}(Z_{IG})$. The results are shown in Fig. \ref{fig7}, where the predicted values of $\Str$  given by the three above mentioned criteria are plotted as function of $\Blc$. Beginning the discussion with the results obtained for the wake without base bleed, $\Blc=0$, it can be observed that the global Strouhal numbers predicted by the IG and HR criteria are very close to each other, namely $0.23$ and $0.22$ respectively, while the MG criterion provides a much lower value of $0.09$. The effect of an increasing base bleed on the Strouhal number predicted by the IG criterion is
non-monotonic: it first decreases, until a minimum value of $0.14$ is reached at $\Blc\approx 0.065$; for higher values of $\Blc$ the Strouhal number increases again up to a value of $0.19$ near the critical bleed coefficient $\Blc^*\approx 0.125$. In the case of the MG criterion, $\Str$ increases monotonically with $\Blc$ until it reaches a value of $0.18$ near criticality. Finally, the HR criterion predicts an initial decrease to a value of $0.18$, followed by a plateau up to the critical blowing. Notice that the frequency predictions based on the three criteria tend to the same value as $\Blc\rightarrow\Blc^*$, since the AGR vs. $Z$ curve becomes tangent to the $Z$ axis at this point.

\section{Experimental results}

\begin{figure}[!h]
\begin{center}\includegraphics[width=0.6\textwidth]{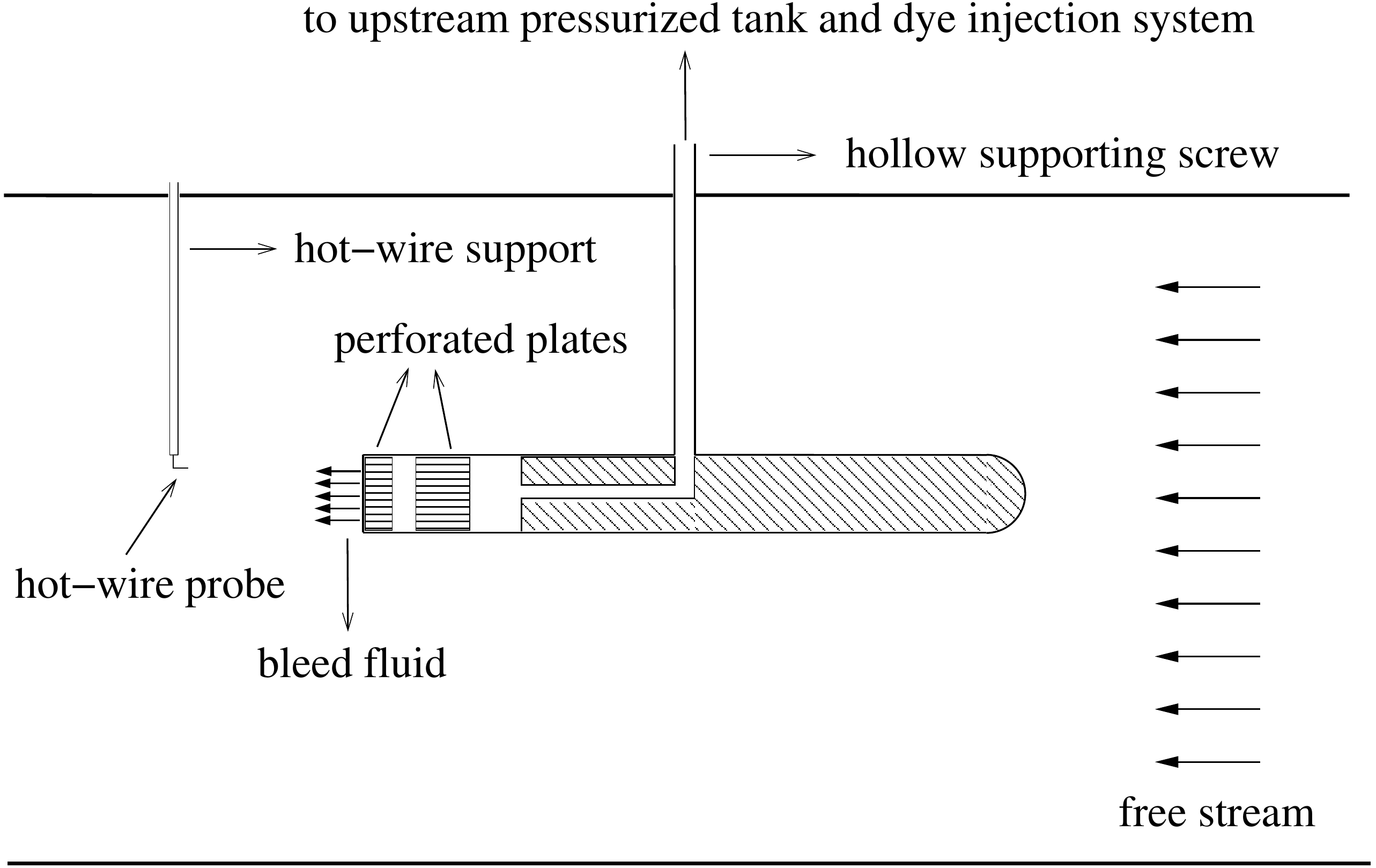}\end{center}
\caption{Sketch of the experimental set-up.}\label{fig8}
\end{figure}

The experimental set-up used in the present work is depicted in Fig.~\ref{fig8}. The axisymmetric object under study has a 3:1 aspect ratio semi-ellipsoidal nose carefully attached to a cylindrical body  with a blunt trailing edge. The diameter of the body is $D\approx 17.1\;\text{mm}$, and its length is $\,l\approx 168\;\text{mm}$, giving a total length to diameter ratio of $\text{L}\approx 9.8$. Flow visualizations were performed in a recirculating water channel, and hot-wire measurements were made in a low-speed wind tunnel. The uniformity of the flow was ensured in both facilities by previous flow characterization studies. The bleed fluid was supplied from an upstream pressurized tank, and the bleed flow rate was regulated and measured by flow meters provided with high resolution valves. The bleed fluid passed through the supporting rod, the hollow core of the body, and finally through a pair of perforated plates situated near the base of the model (see Fig.~\ref{fig8}).

In the next two subsections we will present the results of flow visualizations in the water channel as well as hot-wire measurements performed in the wind tunnel.

\subsection{Flow visualizations}

The range of free-stream velocities used in the water channel was $3\times 10^{-2}\,\text{ms}^{-1}<u_\infty<1.8\times 10^{-1}\,\text{ms}^{-1}$, which corresponded to Reynolds numbers from $510$ to $3000$ approximately. Flow visualizations were performed by means of controlled dye injection from a pressurized mixing chamber placed in series with the bleed line, and the bleed flow rate was measured with a flow meter. Video recordings were taken with a high speed camera whose frame rate was selected in all cases to be higher than three times the vortex shedding frequency, estimated as $f\approx 0.26\,u_\infty/D$.

\begin{figure}[!h]
\begin{center}\includegraphics[width=0.5\textwidth]{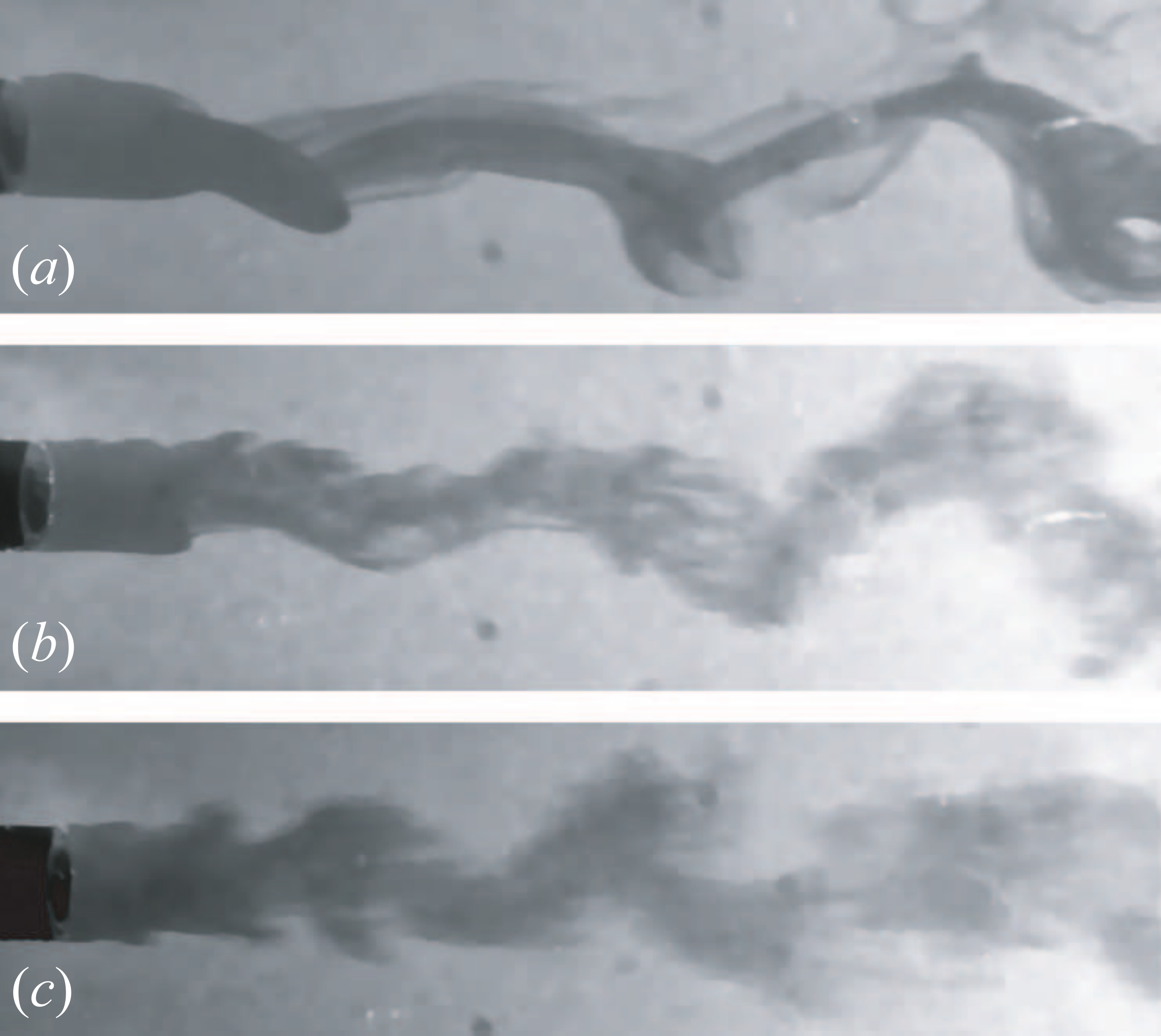}\end{center}
\caption{Visualizations of the wake without base bleed for several values of the Reynolds number: (a) $\Rey= 1075$, (b) $\Rey=1900$ and (c) $\Rey=2650$.}\label{fig9}
\end{figure}

Several photographs of vortex shedding for the natural wake, $\Blc\approx 0$, are shown in Figs.~\ref{fig9}(a)--(c) where the Reynolds number takes the values $1075,\,1900\,\text{and}\,2650$ from top to bottom. It can be observed that the wake is almost laminar in the first photograph, \ref{fig9}(a), with periodic release of hairpin vortex structures taking place. In Figs.~\ref{fig9}(b)--(c) the appearance of smaller structures is clearly seen, indicating transition to turbulence in the wake. The vortex shedding phenomenon is still observed as large-scale structure, and its spatial characteristics resemble those previously found for turbulent wakes of other axisymmetric bodies like spheres, cones and disks. During the visualizations it could be observed how the wake performed a sinuous oscillation in a horizontal plane while keeping symmetry with respect to the perpendicular plane. According to Monkewitz~\cite{monk88c}, this axisymmetry-breaking instability corresponds to the superposition of the $m=\pm 1$ azimuthal instability modes, which were found to be absolutely unstable in a bounded region of the near wake. It has to be pointed out, however, that the spatiotemporal coherence of the shed structures in the turbulent regime is much smaller than in the laminar shedding regime, and eventually disappears in the far field of the wake.

\begin{figure}[!h]
\begin{center}\includegraphics[width=0.5\textwidth]{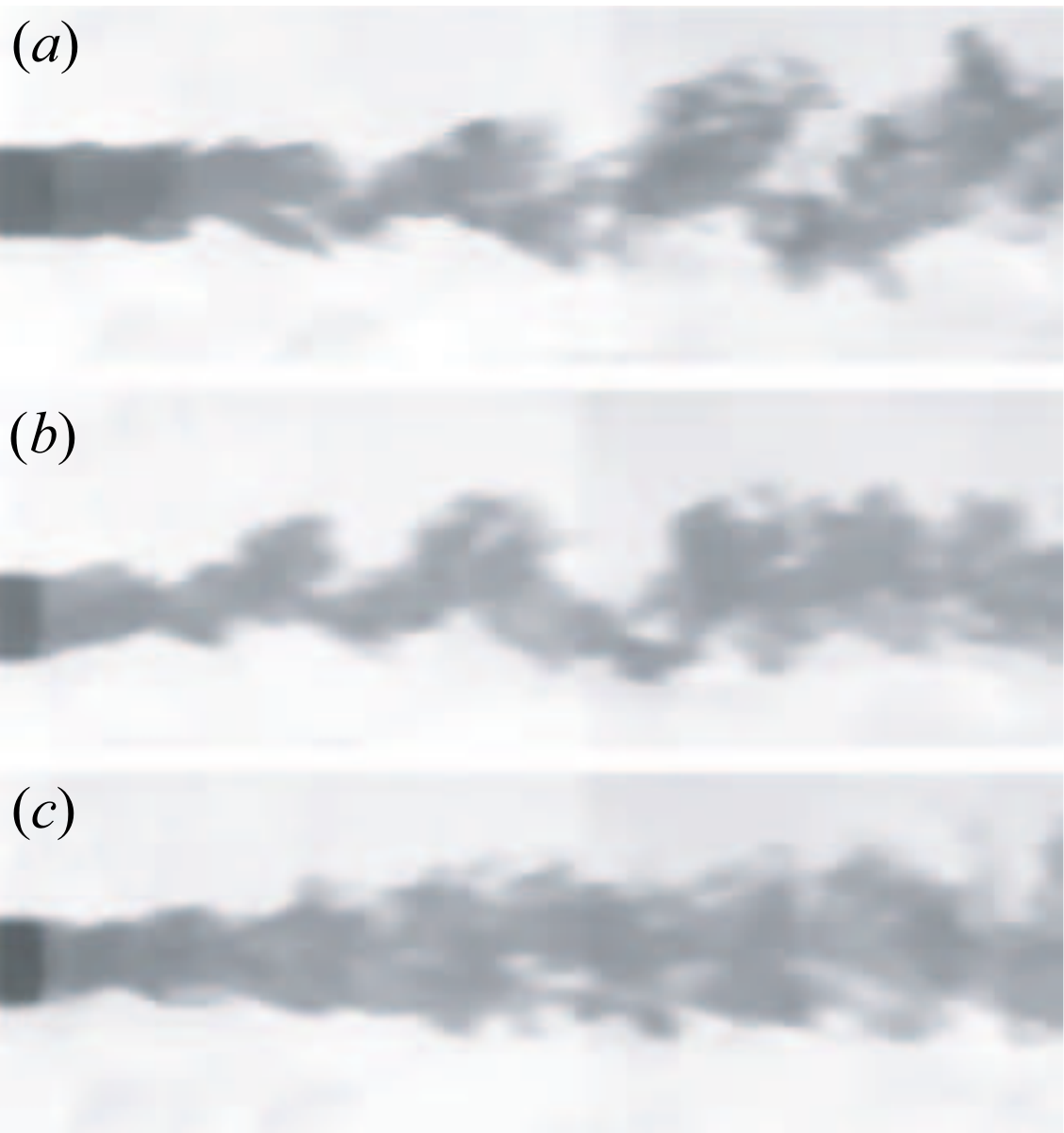}\end{center}
\caption{Visualizations of the wake for $\text{Re}=2800$ and several values of the bleed coefficient: (a) $\Blc= 0.04$, (b) $\Blc=0.09$ and (c) $\Blc=0.13$.}\label{fig10}
\end{figure}

Figure \ref{fig10} shows some images of the wake for increasing values of the base bleed from top to bottom and a fixed value of the Reynolds number, $\Rey\approx 2800$. Notice that figures \ref{fig10}(a)--(b), where the values of the bleed coefficient are $\Blc=0.04\,\text{(a)},\,0.09\,\text{(b)}$, clearly show large-scale vortex shedding in the near wake field. However, in the last photograph \ref{fig10}(c), where $\Blc=0.13$, large-scale coherence is inhibited and the mean wake flow becomes axisymmetric. Also note the remarkable agreement between the experimental value of the critical bleed coefficient, $C_{b,\,\text{EXP}}^*\approx 0.13$, and the value obtained with the linear instability analysis presented in section II, $0.125<C_{b,\,\text{LI}}^*<0.14$.

\subsection{Hot-wire measurements}
Hot wire measurements were performed in a wind tunnel operated in the range of velocities $3.7\,\text{ms}^{-1}<u_\infty<9.1\,\text{ms}^{-1}$, giving a Reynolds number range $4140<\Rey<10273$.  Axial-velocity signals were obtained with a hot wire anemometer. The hot wire probe was located at $\eta\approx 0$ and $Z\approx 1.5$, and the sampling rate was in all cases $2048\,\text{s}^{-1}$. The shedding frequency was determined by performing a spectral analysis of the raw data.

\begin{figure}[!h]
\begin{center}\includegraphics[width=0.45\textwidth]{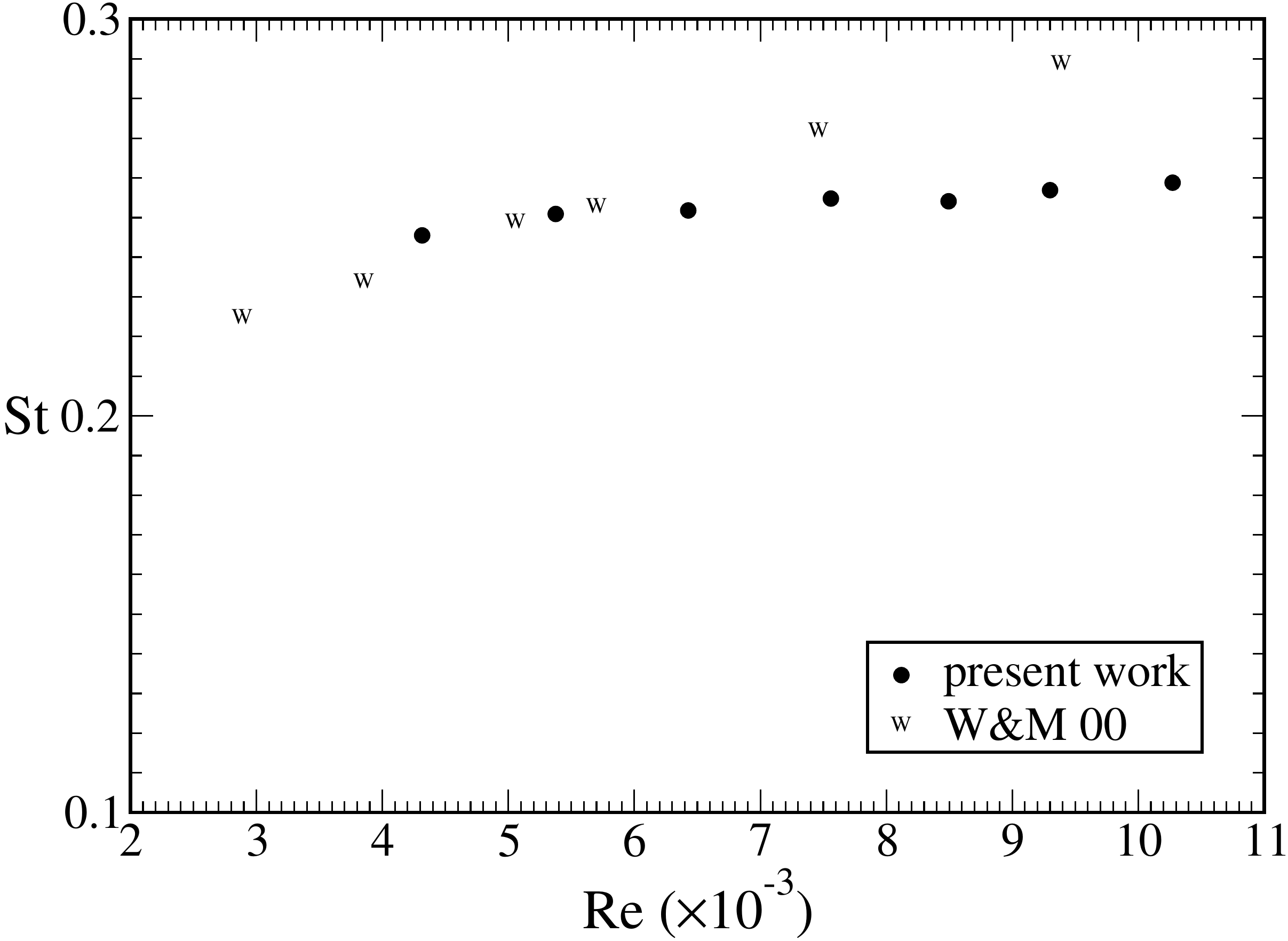}\end{center}
\caption{Evolution of  the Strouhal number $\text{St}$ with the Reynolds number $\text{Re}$ for $\text{C}_b=0$. The Strouhal number is based on hot-wire measurements at $Z=1.5$. Measurements performed by Weickgenannt and Monkewitz~\cite{weick00}, W\&M00, have been included for comparison.}\label{fig11}
\end{figure}

Figure \ref{fig11} shows the dependence of the Strouhal number on the Reynolds number. Unfortunately, hot-wire measurements in the wind tunnel could only be performed for values of the Reynolds number higher than $4000$, so we decided to include also the results of Weickgenannt and Monkewitz\cite{weick00} --hereafter W\&M00-- in Fig.~\ref{fig11} for comparison. The difference between the control mechanism of W\&M00 and that explored here is that they used a rear-mounted control disk attached to the body.In W\&M00 the main control parameter is the non-dimensional distance between the disk and the base, $s$. The results of W\&M00  plotted in Fig.~\ref{fig11} of the present work correspond to a value $s=0.28$, which is the lowest one available in Fig. $4$ of W\&M00. It can be observed in Fig.~\ref{fig11} that both results collapse, showing an increase of the Strouhal number with the Reynolds number for $\Rey < 6\times 10^3$. However, for higher values of the Reynolds number the two curves exhibit different behavior: while the Strouhal number obtained in W\&M00 continues to increase appreciably as the Reynolds number increases, our results show a constant value in the range $0.25\lesssim\Str\lesssim 0.26$ for $\Rey >  6\times 10^3$. Such difference may be attributed to an increasing influence of the rear-mounted disk present in W\&M00 as the Reynolds number increases, as can also be deduced from Fig.~ $4$ of W\&M00.

A comparison of the experimental value of the Strouhal number for the natural wake with the predictions made by linear instability is now possible. The IG, MG and HR criteria give the values $\Str_{\text{IG}}=0.23,\;\Str_{\text{MG}}=0.09,\;\Str_{\text{HR}}=0.22$ respectively (see Fig.~\ref{fig7}) when $\Blc=0$, while the experimental value for $\Rey\approx 3\times 10^3$ is $\Str_{\text{EXP}}\approx 0.225$. Thus, we can conclude that both the IG and HR criteria give good predictions for the Strouhal number, while the MG criterion considerably underpredicts the shedding frequency. It is important to point out that the success of the IG criterion in predicting the global frequency is, in fact, a success of the theory of steep non-linear global modes, which seems to be useful even when the hypothesis of weak non-parallelism is not satisfied.

\begin{figure}[!h]
\begin{center}\includegraphics[width=0.5\textwidth]{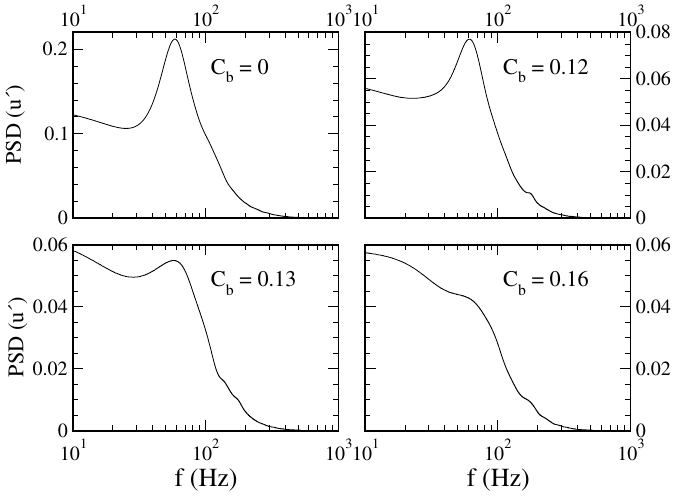}\end{center}
\caption{Spectra of axial-velocity hot-wire signals at $Z=1.5,\,\eta=0$ for $\text{Re}=4200$ and increasing values of the bleed coefficient.}\label{fig12}
\end{figure}

\begin{figure}[!h]
\begin{center}\includegraphics[width=0.5\textwidth]{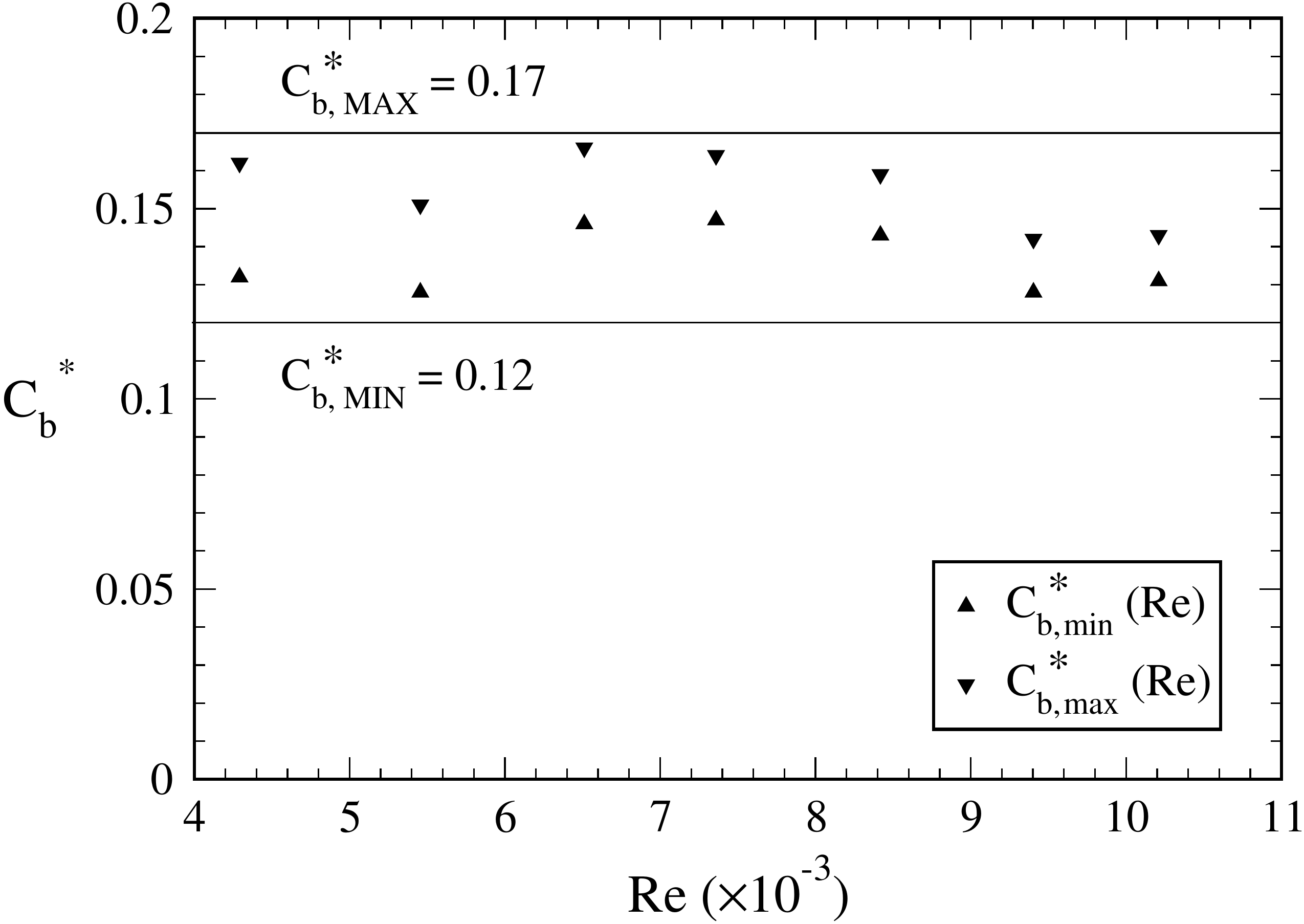}\end{center}
\caption{Experimental results of the critical bleed coefficient, $\text{C}_b^*$ as a function of the Reynolds number.}\label{fig13}
\end{figure}

In addition to studying the natural wake with the hot-wire technique, we also performed a series of measurements increasing the  bleed coefficient until the critical value was achieved at a fixed Reynolds number. An example of the results obtained is shown in Fig.~\ref{fig12}, where the power spectral density associated with axial-velocity hot-wire signals is plotted for $\Rey=4200$ and four increasing values of the bleed coefficient, $\Blc=0,\,0.12,\,0.13,\,0.16$. As already confirmed by flow visualizations, the spectral peak observed at $f\approx 60\,\text{Hz}$ corresponds to large-scale vortex shedding phenomenon. Notice that the magnitude of the spectral peak decrease as $\Blc$ increases until eventually disappears for the last value of the bleed coefficient. Thus, we can conclude that for this value of the Reynolds number the critical bleed coefficient is bounded between a minimum value of $\text{C}_{b,\;\text{min}}^*=0.13$ and a maximum one of $\text{C}_{b,\;\text{max}}^*=0.16$. Following the same routine  we measured the critical bleed coefficient for different values of the Reynolds number. The results obtained are shown in Fig.~\ref{fig13}, where the minimum and maximum values of the critical bleed coefficient are plotted as a function of Reynolds number. In this figure we have also included the overall minimum value $\text{C}_{b,\,\text{MIN}}^*\approx 0.12$, and maximum value, $\text{C}_{b,\,\text{MAX}}^*\approx 0.17$, for the whole range of Reynolds numbers studied. Notice that the dependence of $\text{C}_{b}^*$ on the Reynolds number is very weak, and that the value predicted by linear instability, $\text{C}_b^*\approx 0.13$, falls within the experimental uncertainty. 

The remarkably good agreement between the experiments and linear instability predictions also confirms Monkewitz's hypothesis~\cite{monk88c} suggesting that the basic mechanism underlying helical vortex shedding in axisymmetric wakes is a region of absolute instability in the near field for the $m=\pm 1$ azimuthal mode. Under this point of view, the effect of base bleed can be summarized as follows. Base bleed acts on the basic velocity profiles of the near wake field by reducing the shear. In the local instability approach, this causes the first azimuthal mode to become locally convectively unstable everywhere for sufficiently high values of the bleed coefficient and, consequently, the non-linear global mode which is responsible of vortex shedding can no longer survive.

\section{Conclusions}
In the present investigation we have studied large-scale helical vortex shedding in the wake of an axisymmetric body with a blunt trailing edge, and its global control by means of base bleed. To perform the study we have combined a local, linear stability analysis of the near wake field with both flow visualizations and hot-wire measurements.

It has been observed that a simple analysis based on local, linear, spatiotemporal instability can lead to good global predictions with a very low computational cost compared to three-dimensional direct numerical simulations. We have found that the use of local analysis, with the aid of both the initial growth (steep global) and hydrodynamic resonance criteria, provides a value for the natural-wake shedding Strouhal number which agrees remarkably well with the corresponding experimental value. On the contrary, the maximum growth criterion considerably underpredicts the shedding frequency of the natural wake. The success of the initial growth criterion can be understood taking into account recent studies on non-linear stability theory. Quoting Pier \& Huerre~\cite{pier01},\\

``\emph{It is somewhat paradoxical that the intricate complex $X$-plane analyses developed over the years to uncover the selection properties of linear global modes are masked in practice by the onset of local absolute instability which immediately prevails and imposes its frequency and the overall structure of the synchronized oscillations.}''

Thus, the IG criterion coincides with the selection criterion for steep global modes and, since the latter seem to be responsible for the vortex shedding phenomenon, the agreement we found using the IG criterion is in fact a success of the non-linear global mode theory. However, it is important to emphasize the fact that these theories have been developed for the case of slowly divergent flows, while in our case this requirement is not accomplished for low values of the bleed coefficient. A rigourous approach to the problem would consist of a global linear analysis of the two-dimensional flow described by Eqs.~\eqref{conservation2}, an investigation which should be addressed in the future.

Furthermore, inhibition of the instability has been explored by means of base bleed. The critical value of the bleed coefficient for inhibition of large scale vortex shedding, $\text{C}_b^*$, has been obtained both experimentally and from instability analysis. The theoretical value, $\text{C}_{b,\,\text{LI}}^*\approx 0.13$ for $\Rey =3\times 10^3$, has been based on the fact that a sufficient bleed flow rate leads to a locally convectively unstable wake at all downstream positions. In addition, minimum and maximum experimental limits for the critical bleed coefficient, $0.12<\text{C}_{b,\,\text{EXP}}<0.17$, have been obtained over the Reynolds number range $4\times 10^3<\Rey<1.2\times 10^4$, showing that the value predicted by linear instability falls within the experimental uncertainty.

\begin{acknowledgments}
This work has been supported by the Spanish MCyT under project \# DPI2002-04550-C07-06. The authors wish to thank Dr. J. M. Gordillo for helpful discussions.
\end{acknowledgments}


\end{document}